\documentclass[12pt]{article} 
\usepackage{amsmath,amsfonts,amssymb}
\usepackage{latexsym}
\usepackage{epsfig}
\usepackage{textcomp}

\newcommand{\be}{\begin{equation}}
\newcommand{\en}{\end{equation}}

\newcommand{\ii}{\textrm{i}}
\newcommand{\ee}{\textrm{e}}
\renewcommand{\vec}[1]{\boldsymbol{#1}}

\begin{document}

\title{Compact travelling waves in viscoelastic solids}

\author{Michel Destrade$^a$, Pedro M. Jordan$^b$, Giuseppe Saccomandi$^c$\\[12pt]
$^a$School of Electrical, Electronic, and Mechanical Engineering,\\
University College Dublin, Belfield, Dublin 4, Ireland\\[12pt]
$^b$Code 7181, Naval Research Laboratory, \\
Stennis Space Center,
 MS 39529--5004, USA\\[12pt]
$^c$Dipartimento di Ingegneria Industriale, \\
Universit\'a degli Studi di Perugia, 06125 Perugia, Italy}

\date{}

\maketitle
\bigskip%

\begin{abstract}

We introduce a model for nonlinear viscoelastic solids where traveling shear waves with compact support are possible. 
We obtain an exact compact solution. 
We also derive a new Burger's type evolution equation associated with the introduced constitutive equation.

\end{abstract}

\bigskip

\emph{PACS:} 

94.05.Fg (Solitons and solitary waves), 

43.35.Mr (Acoustics of viscoelastic materials), 

46.00.00 (Continuum mechanics of solids).

\newpage

 
\section{Introduction}

 
A \emph{compact wave} is a non-linear solitary wave  with a definite amplitude, which exists within the confines of a compact support; outside that support, it vanishes identically. 
Compact waves  can be used to describe patterns with a compact support and sharp fronts.
These are ubiquitous in Nature, but are hard to model mathematically. The \emph{driving force} for the modeling of compact waves has been the 1993 discovery of compactons by 
Hyman and Rosenau~\cite{HR}. A \emph{compacton}, in full analogy with the definition of a soliton, is a compact wave which preserves its shape and amplitude after a collision with another compact wave.

Compact waves emerge essentially by mathematical degenerancy of the equations of motion, leading to a local loss of uniqueness, which enables the patching of two different solutions with a certain regularity. 
This point may be clarified and made rigorous in several ways, see Saccomandi \cite{Sacco} and Destrade \emph{et al.} \cite{Destrade} who use the classical Weierstrass criterion to generate compact waves, and also Gaeta \emph{et al.} \cite{Gaeta}, for example.

From a purely mathematical point of view, there exist several equations that are \emph{factories} of compact waves, such as for example the $K(m,n)$ KdV equation \cite{HR}. 
On the other hand, there exist very few examples of physically-based equations which are capable of generating compact waves within the framework of a rigorous theory of material behavior. 
For solids, three examples of such rigorous derivations have been presented so far: \emph{(i)} in 1998 Dusuel \emph{et al.} \cite{Remo} show that, in the continuum limit, the generalized $\Phi$-four or double-well model with nonlinear coupling can exhibit compacton-like kink solutions when the nonlinear coupling between pendulums is dominant; using these ideas Saccomandi and Sgura \cite{Saccomandi} give an analysis of the full non-linear version of the classical models of DNA; 
\emph{(ii)} Destrade and Saccomandi \cite{Destrade1} use a non-linear theory of dispersion  compatible with the axiomatic foundation of simple materials; 
\emph{(iii)} Goriely \emph{et al.} \cite{Goriely} extend the theory of rods to non-linear material laws. In fluids the emergence o compact coherent structures, mainly via asymptotic method, has been a more investigated topic \cite{fluids}.

Here, we show that it is possible to generate compact waves in the framework of the nonlinear theory of viscoelasticity.  
To this end we consider a special viscoelastic theory of incompressible isotropic solids for which the viscous part is that of the Navier-Stokes theory, with a shear dependent viscosity---a common assumption in non-Newtonian fluid mechanics.
 
In this framework, for a special class of constitutive equations, we provide a rigorous existence result, an exact (albeit implicit) exact solution and an approximation for the compact kink. Moreover, we derive a new generalized Burger's equation as asymptotic  reduction of the full equation for the propagation of shear waves.

 
\section{Governing equations}

We call $\vec{X}$ the position of a particle in the solid in $\left( \mathcal{B}_r\right)$, the reference configuration, and $\vec{x}$ the position of that particle at time $t$ in $\left( \mathcal{B}\right)$, the current configuration. 
A motion of the body is the one-to-one mapping $\vec{\chi}$  such that $ \vec{x}=\vec{\chi}(\vec{X},t)$.  
The deformation gradient $\vec{F}$ and the left Cauchy-Green tensor $\vec{B}$ associated with this motion are
\be
\vec{F}=\frac{\partial \vec{\chi}}{\partial \vec{X}},
\qquad 
\vec{B=FF}^{T},  \label{05}
\en
respectively, and the strain-rate tensor is defined as $\vec{D}= \tfrac{1}{2}\left(\dot{\vec{F}}\vec{F}^{-1}+\vec{F}^{-T}\dot{\vec{F}}^T \right)$, where the superposed dot denotes the material time derivative.
An incompressible solid can undergo only isochoric motions, and this internal constraint translates mathematically as: $\det \vec{F}=1$, $\text{tr }\vec{D}=0$, at all times. 

We are interested in viscoelastic materials of differential type, with Cauchy stress tensor
\begin{equation}
\vec{T}=-p\vec{I} + 2 \beta_1 \vec{B}-2 \beta_{-1} \vec{B}^{-1} + 2\hat\nu \vec{D}.
\label{1}
\end{equation}
Here, $p$ is the indeterminate Lagrange multiplier introduced by the incompressibility constraint, $\beta_1$ and $\beta_{-1}$ are the elastic response parameters, and 
$\nu$ is the shear viscosity coefficient.
In all generality, $\beta_i = \beta_i (I_1, I_2)$ ($i=-1,1)$, where $I_{1}$, $I_{2}$ are the first two principal invariants of the Cauchy--Green strain: 
$I_{1}=\mathrm{tr}(\vec{B})$ and $I_{2}=\text{tr}(\vec{B}^{-1})$.
We assume that $\hat\nu=\hat\nu \left(\vec{D \cdot D}\right)$, and moreover, that $\hat\nu>0$, i.e. 
the model is dissipative.

The momentum equations, in the absence of body forces, take the form
$\text{div}\; \vec{T} = \rho \partial\vec{v}/\partial t$,
where $\rho$ is the mass density and $\vec{v} = \partial \vec{\chi}/\partial t$ is the velocity.
Our aim is to investigate what happens in the \emph{shearing motion}, specifically, $x=X+\mathfrak{f}(Z,t), \, y=Y, \,z=Z,$ 
where the function $\mathfrak{f}$ is as yet unknown. Straightforward computations give the components as
\be \nonumber 
\left( B_{ij}\right) =
\left(
\begin{array}{ccc}
1+ K^{2} & 0 & K
\\
0 & 1 & 0
\\
K & 0 & 1
\end{array}
\right),  
\en
and
\be \nonumber 
2\left(D_{ij}\right) =
\left(
\begin{array}{ccc}
0 & 0 & K_{t} 
\\
0 & 0 & 0
\\
K_{t} & 0 & 0
\end{array} 
\right),
\en
where $K \equiv \mathfrak{f}_{Z}$ is the \emph{ amount of shear}, and the subscript denotes partial differentiation. Clearly we have now $I_1 = I_2=3+K^2$ and $\vec{D \cdot D}=K_{t}^2/2$.

With $p_x \equiv 0$, two of the three equations of motion are identically satisfied. 
The remaining equation is $\rho \mathfrak{f}_{tt}=\partial T_{13}/\partial Z$. 
Therefore the determining equation for the amount of shear $K$ becomes
\be \label{fe}
\rho K_{tt}=\left[\hat Q(K^2)K+\hat\nu(K_t^2) K_{t} \right]_{ZZ},
\en 
where  $\hat Q \equiv 2(\beta_1+ \beta_{-1})$ is the \emph{generalized shear modulus}. The mathematical theory 
of quasilinear equations for viscoelasticity of strain-rate type can be found in ref.~\cite{Tvedt}. 

In order to rewrite eq.~\eqref{fe} in a dimensionless form, we need a characteristic frequency $\Omega(>0)$ so that it is possible to introduce the dimensionless time $\tau=\Omega t$; 
we also need a characteristic length $L$ so that we can introduce the dimensionless length $\zeta=Z/L$. 
Usually the length $L$ is determined by the geometry of the problem (e.g. the thickness of a slab wherein the wave is propagating). 
The characteristic frequency $\Omega$ may be introduced in several ways: via the boundary conditions, by defining the ratio $\Omega=\hat\mu_0/\hat\nu_0$, where $\hat\mu_0=\lim_{K^2 \rightarrow 0} \hat Q$ is the infinitesimal shear modulus and $\hat\nu_0=\lim_{K_t \rightarrow 0} \hat\nu$, or by a characteristic (finite) time $t^*$ at which, for example, localization of the solution occurs. 
Eventually, eq.~\eqref{fe} becomes 
\be \label{fed}
\delta K_{\tau \tau}= [QK+\nu K_{\tau}]_{\zeta \zeta},
\en 
where $\delta=\rho \Omega^2/\hat \mu_0 L^2$,  $Q=\hat{Q}/\hat \mu_0$, and $\nu=\Omega\hat{\nu}/\hat\mu_0$.

As the final step, we specialize the equation of motion  (via the constitutive relations) to the case of fourth-order elasticity. In particular, we take $ Q=1+\mu_1 K^2$, where $\mu_1$ is a constant ($\mu_1>0$ for strain-stiffening solids and $\mu_1<0$ for strain-softening solids), and assume the simplest form of shear viscosity dependence, namely, $\nu =\nu_0+\nu_1 K_{\tau}^2$, where $\nu_1>0$ is a constant. Note that $\mu_1 = (\mu + A/2 + D)/\mu$, where $\mu$, $A$,  and $D$ are
the second-, third-, and fourth-order constants of weakly non-linear
elasticity \cite{Ogde74, JASA04}. Under these assumptions, eq.~\eqref{fed} reduces to 
\be \label{feds}
\delta K_{\tau \tau} = \left[K+\mu_1 K^3+\nu_0 K_{\tau}+\nu_{1}K^3_{\tau} \right]_{\zeta \zeta}.
\en 
The existence and regularity of solutions for the Cauchy problem of such kind of equations has been considered by Friedman and Necas \cite{Friedman}.  
Also, Pucci and Saccomandi \cite{Pucci} considered the quasistatic limit of eq.~\eqref{feds} and studied the mathematical and mechanical properties of the  classical creep and recovery experiment.

Eq.~\eqref{feds} is neither integrable nor linearizable, as may be checked via symmetry
arguments \cite{Valenti} so that we do not expect our solitary
compact waves to preserve their shape and amplitude after colliding. In all
likelihood, they are not compactons.


\section{Kinks}


In this section, we seek travelling wave solutions (TWS)s of eq.~\eqref{feds} in the form of kinks; i.e., continuous, bounded, monotonic waveforms that tend to constant, but unequal, limits at $\pm\infty$.
It is known that \emph{kinks} may propagate in a viscoelastic medium; 
Jordan and Puri \cite{Jordan1,Jordan2} give an explicit characterization of such waves, and a detailed survey of the various qualitative properties of travelling waves solutions in viscoelasticity may be found in~\cite{Antman}.

We begin our search for TWSs with the following observe: since eq.~\eqref{feds} is  invariant under the transformation $\zeta \mapsto -\zeta$,  we need only consider, without loss of generality, right-traveling waves, i.e., solutions of the specific form $K(\zeta,\tau)=  g(\xi)$, where  $g$ is a function of the single variable $\xi \equiv \zeta - c t$ and the positive constant $c$ denotes the wave speed. 
Substitution of the travelling wave ansatz into eq.~\eqref{feds} results in the following nonlinear ordinary differential equation (ODE):
\be \label{feds1}
(1-\delta c^2) g''+\mu_1 (g^3)''-\nu_0cg'''- \nu_{1}c^3 [(g')^3]''=0,
\en 
where primes denote differentiation with respect to $\xi$. Integrating this ODE twice, setting the first integration constant to zero, and then enforcing the usual (kink) asymptotic conditions\footnote{That is, $g \rightarrow g_{1,2}$, as $\xi \rightarrow \mp \infty$, where $g_1 > g_2 \geq 0$ are constants.}  yields 
\be \label{feds2bis}
(g')^3 + \tilde \nu g' = \frac{(1-\delta c^2) g + \mu_1(g^3 + g_1^2 g_2 + g_2^2 g_1)}{\nu_1c^3}.
\en 
Here, $\tilde \nu = c^{-2}\nu_0/\nu_1$ and the wave speed  is given by 
\be \label{c}
c=\delta^{-1/2}\sqrt{1+\mu_1(g_{1}^2+g_1 g_2+g_{2}^2)}.
\en

Having obtained the first order ODE satisfied by our TWS and determined the wave speed, it is instructive to now consider the  cases $\tilde\nu=0$ and $\tilde\nu>0$ separately.
\subsection{The case $\nu_0=0$}
 Focusing  on this, the simplest case first, we  further simplify the analysis by taking $g_2=0$ and setting $f=g/g_1$; in particular,
$c$ reduces to $c_1$, where
\be
c_1= \sqrt{(1+\mu_1g_{1}^2)/\delta},
\en
and eq.~\eqref{feds2bis} becomes 
 \be \label{normal}
 \left( f' \right)^3 = -\sigma f (1-f^2), \qquad
 f \in [0,1],
 \en
where we have set $\sigma = c_{1}^{-3}\mu_1/\nu_1$ for convenience.  
Now a standard stability analysis of eq.~\eqref{normal} 
reveals that $f=0, 1$, the equilibrium solutions relevant to our investigation,  are
stable and unstable, respectively, for $\sigma>0$. 
Henceforth limiting our attention to only those dispersive solids that stiffen in shear 
(i.e., those for which $\mu_1 >0$), we separate variables in eq.~\eqref{normal} and integrate. We are thus led to consider  the quadrature 
\be \label{integral}
\int \frac{{\rm d}f}{[f(1-f^2)]^{1/3}}=-\sigma \xi+\mathcal{K},
\en
where the integration constant $\mathcal{K}$ will be chosen so that the kink is centered at $f(0)=1/2$. 
Because of the zeros at $f=0$ and $f=1$ in the  denominator of the integrand, the left-hand side of eq.~\eqref{integral} is, in fact, a generalized integral (see~\cite{Destrade}). 
Nevertheless, it can be evaluated exactly in terms of special functions. Omitting the detail, it is readily established that the exact, albeit implicit, solution is given by
\be \label{exactsol}
_2F_1(\tfrac{1}{3},\tfrac{1}{3};\tfrac{4}{3};
f^2)f^{2/3}-\,_2F_1(\tfrac{1}{3},\tfrac{1}{3};\tfrac{4}{3};
\tfrac{1}{4})(\tfrac{1}{4})^{1/3}=-\tfrac{2}{3}\sigma^{1/3} \xi, \quad
\en
for $\xi\in (\xi_1,\xi_0)$, while outside this interval  we have
\be \label{out}
f(\xi)=\begin{cases}
1, &\xi\leq \xi_1,\\
0, &\xi\geq \xi_0.
\end{cases}
\en
Here, $_2F_1$ denotes the Gauss hypergeometric series and the constants $\xi_1$, $\xi_0$ are defined by
\be
 \xi_1=\frac{-2\pi\sqrt{3}+3B_{1/4}(
\tfrac{1}{3},\tfrac{2}{3})}{6\sigma^{1/3}},
\quad
\xi_0=\tfrac{1}{2}\sigma^{-1/3}B_{1/4}(\tfrac{1}{3},\tfrac{2}{3}),
\en
where
\be
B_{q}(a,b) \equiv \int_{0}^{q}\vartheta^{a-1}(1-\vartheta)^{b-1}{\rm d} \vartheta,
\qquad 
q>0,
\en
denotes the \emph{incomplete beta function}.

The \textit{shock layer} thickness
$\ell$ has the value
\be \label{layer}
\ell=\left[\lim_{\xi \to -\infty}f(\xi)- \lim_{\xi \to
+\infty}f(\xi)\right]/|f'(0)| =2(3\sigma)^{-1/3}.
\en

\renewcommand\thefigure{1}
\begin{figure}
\epsfig{file=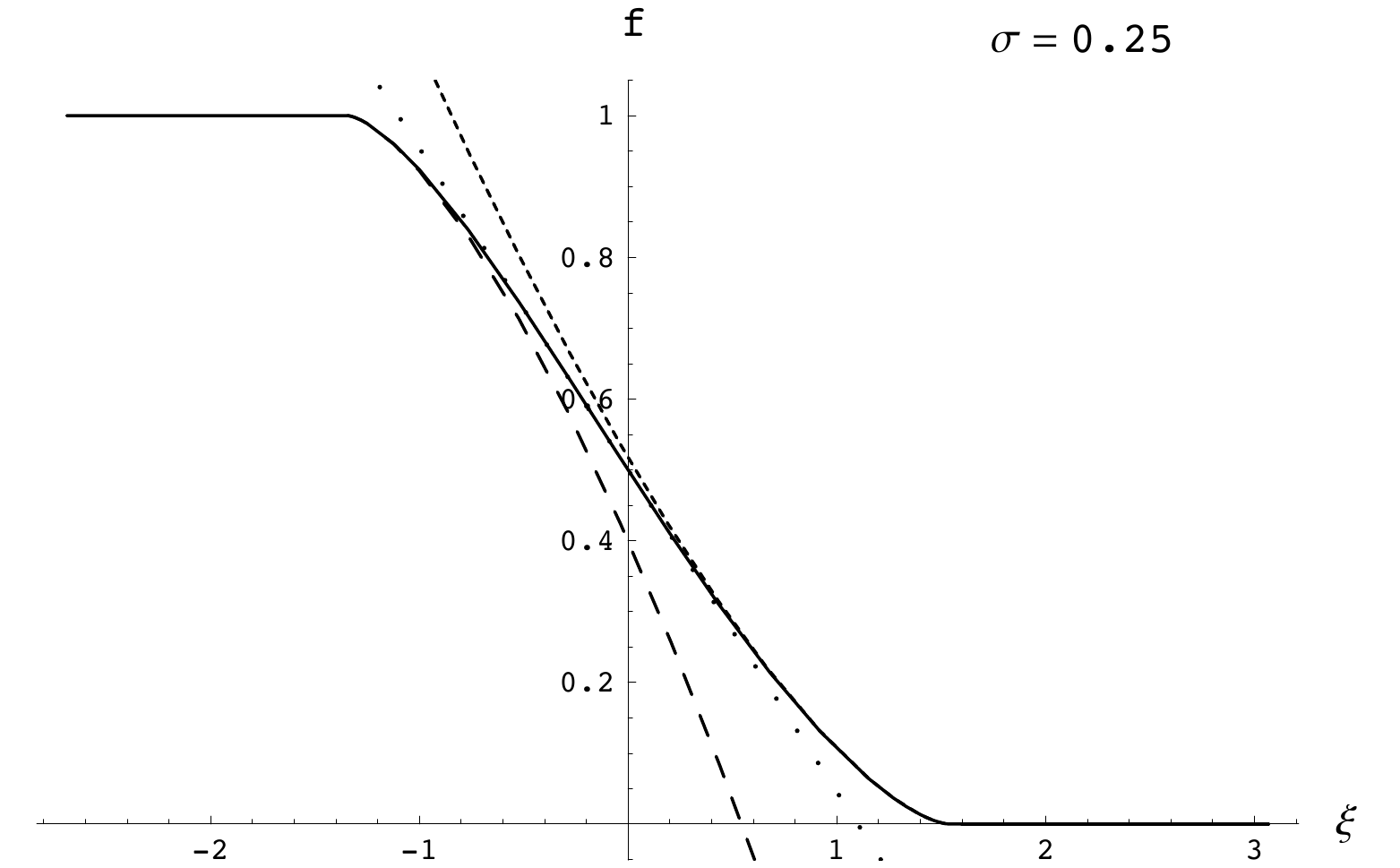, width=75mm}
\caption{$f$ vs.\ $\xi$ for  $\sigma=0.25$ and $\tilde \nu=0$. Solid curve: eq.~\eqref{exactsol} and
eq.~\eqref{out}. Broken curve (long dashes): eq.~\eqref{approx}$_1$. Dots:
eq.~\eqref{approx}$_2$.  Broken curve (short dashes): eq.~\eqref{approx}$_3$.}
\end{figure}

If we expand the first term on the left-hand side of eq.~\eqref{exactsol}
about $f=1, \tfrac{1}{2}, 0$ and then neglect the appropriate higher order terms, the resulting expressions can be solved for $f$ in terms of $\xi$.  Omitting the details, it is a relatively straightforward task, using the Heaviside unit step function, $H(\cdot)$, to construct the respective approximations (see eq.~\eqref{approx}):
\be\label{approx}
f(\xi) \simeq
\begin{cases} 
H(\xi_1-\xi) + H(\xi-\xi_1)\\
\qquad \qquad \quad \times\left\{1- \frac{\sqrt{2}}{27}\left[2\pi \sqrt{3}-3B_{1/4}(\tfrac{1}{3},\tfrac{2}{3}) + 6\sigma^{1/3} \xi\right]^{3/2} \right\}, &  \xi<-\tfrac{1}{2}\ell; \\
\\
\tfrac{1}{2}(1 - 2\xi/\ell),& |\xi|\ll \tfrac{1}{2}\ell;\\
\\
H(\xi_0 - \xi) [_2F_1(\tfrac{1}{3},\tfrac{1}{3};\tfrac{4}{3}; 
\tfrac{1}{4})(\tfrac{1}{4})^{1/3}-\tfrac{2}{3}\sigma^{1/3} \xi]^{3/2}, & \xi>\tfrac{1}{2}\ell.
\end{cases}
\en
As fig.~1 makes clear, the relatively simple approximate expressions given in  eq.~\eqref{approx} are in very good agreement with the exact kink solution within, and even outside of, their theoretical ranges of validity.

{\bf Remark 1:}  From eq.~\eqref{layer} it is evident that a shock wave, \emph{i.e.}, a propagating jump in $f$, forms as $\sigma \to \infty$ since $\ell\to 0$ in this limit~(see fig.~2). In contrast, our compact kink solution does \emph{not} exhibit acceleration waves, also known as ``weak discontinuities''~\cite[\S89]{LL59},  at $\xi=\xi_{1,0}$ because $f'(\xi)\in C(\mathbb{R})$; however, it should  be noted that $\underset{\xi \in \mathbb{R}}{{\rm max}}|f'| \to \infty$ as $\sigma \to \infty$~(again, see fig.~2).

\renewcommand\thefigure{2(a--c)}
\begin{figure}
\epsfig{file=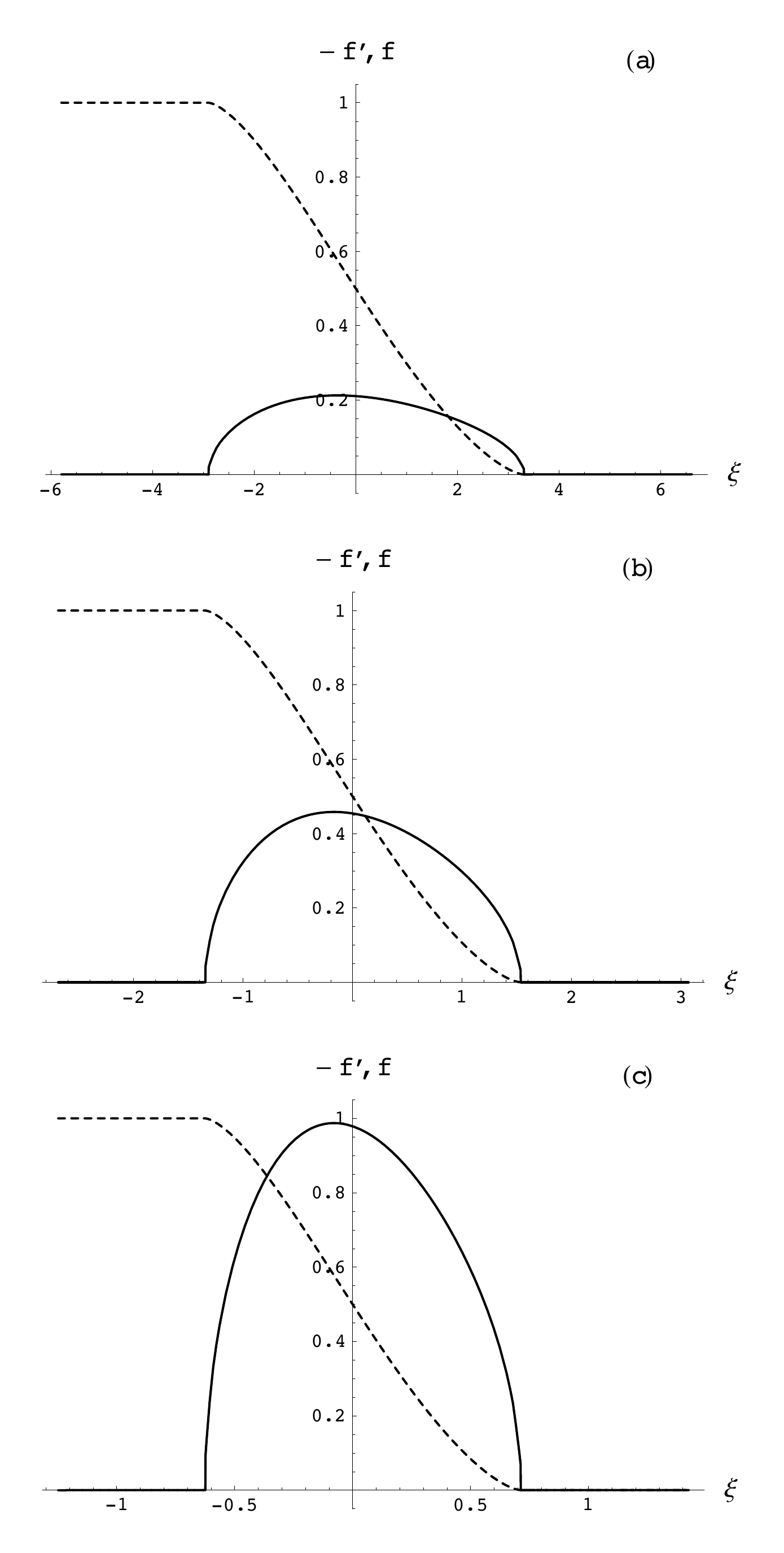, width=75mm}
\vspace*{-5mm}
\caption{Plotted for $\sigma=0.025, 0.25, 2.5$, respectively, and $\tilde \nu=0$.   Solid:~$-f'$ vs.\ $\xi$.
Broken:~$f$ vs.\ $\xi$.}
\end{figure}

\subsection{The case $\nu_0>0$}
Assuming now $\nu_0>0$, we return to eq.~\eqref{feds2bis} and regard this ODE as a cubic polynomial in $g'$.  Using Cardano's formula, the three roots of this cubic, each of which is a function of the single variable $g$,  are readily determined.   Fortunately, however, the cubic discriminant, which we denote here as  $D$, is always positive. Hence, only one of the roots is real-valued. Denoting this particular root by $\mathcal{G}(g)$, eq.~\eqref{feds2bis} becomes
\be \label{fed2tris}
g'=\mathcal{G}(g) \equiv \sqrt[3]{G/2+\sqrt{D}}-\frac{\tilde{\nu}}{3\sqrt[3]{G/2+\sqrt{D}}},
\en
where $D \equiv (\tilde{\nu}/3)^3 + (G/2)^2$ and  $G=G(g)$ denotes the right-hand side of eq.~\eqref{feds2bis}.   We observe here that $\mathcal{G}(g)$ is strictly negative for $g\in (g_2,g_1)$, by Descartes' rule of signs, and that $g^*=\{-(g_1+g_2), g_2, g_1\}$ are the roots of  both $G(g^*)=0$ and $\mathcal{G}(g^*)=0$, where $-(g_1+g_2)$ is an extraneous root in the present context. It should also be noted that the compact kink results discussed in the previous subsection are recovered with little difficulty
by letting $\tilde{\nu} \to 0$ (\emph{i.e.}, $\nu_0\to 0$).

For general values of $\tilde{\nu}$, it is possible to show the existence of a kink solution, but it does  not appear possible to determine its exact analytical  representation.
On the other hand, because $\mathcal{G} = G/\tilde{\nu} - G^3/\tilde{\nu}^4+\cdots$, when expanded about $G=0$,  it is clear that as $\tilde{\nu} \rightarrow 0$, the points where $G = 0$ are associated to a $g'$ which is more and more vertical. 
This means that the kink \textit{compactifies} as $\tilde{\nu}  \to 0$. In other words, as the  nonlinear viscoelastic part in the constitutive function becomes more important with respect to the linear part, the \textit{tails} of the kink are of less importance.   

The process just described is clearly illustrated in fig.~3, where the parameter $\lambda=\tilde{\nu}/g_{1}^{2}$ has been introduced for convenience.  The sequence presented, which was generated from Eq.~\eqref{fed2tris} by once again taking  $g_2=0$ and setting $g=g_1 f$, depicts  the formation of the compact kink travelling wave profile (see Fig.~1) as $\tilde{\nu}  \to 0$. 
\renewcommand\thefigure{3(a--c)}
\begin{figure}
\epsfig{file=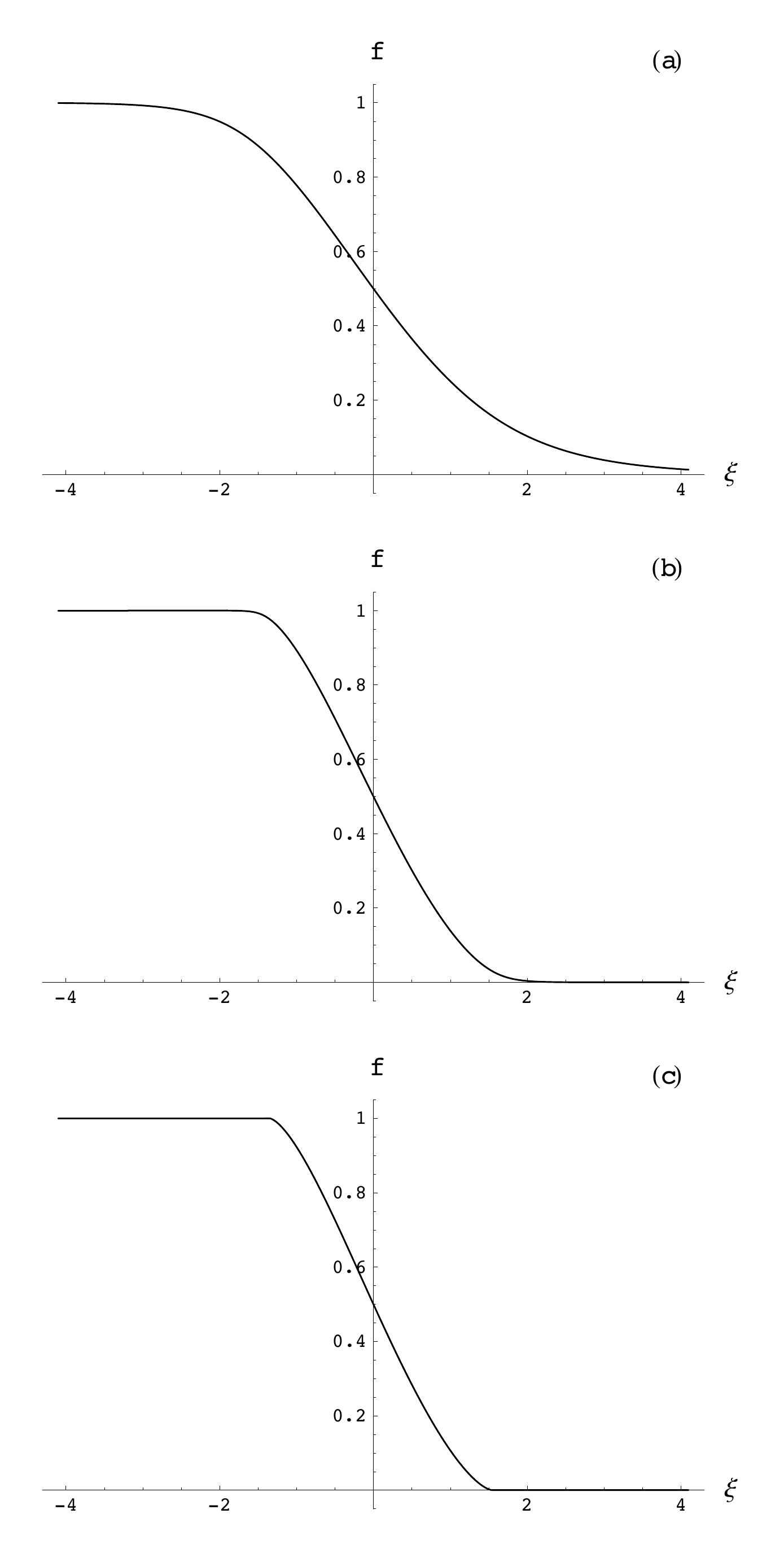, width=75mm}
\vspace*{-5mm}
\caption{$f$ vs.\ $\xi$ for $\lambda=0.25, 0.05, 0.001$, respectively, and $\sigma=0.25$.}
\end{figure}

 Continuing under the assumption $g_2=0$, and with $g=g_1 f$, let us now expand $\mathcal{G}=\mathcal{G}(f)$  about $f=0$. On making the additional assumption $\lambda\gg \sigma^{2/3}$ and neglecting terms $\mathcal{O}(f^4)$, eq.~\eqref{fed2tris} is reduced to the following special case of Abel's equation: $f'\approx -\sigma\lambda^{-1}(f - f^3)$.  Then, taking $f(0)=1/2$, the ``exact'' solution of this ODE is easily found, 
 using~\cite[eqs.~(12)--(18)]{Jordan1}, to be
\be \label{KV_aprox}
f(\xi)\approx  \frac{1}{\sqrt{1+3\exp(2\sigma\xi/\lambda)}} \qquad (\lambda\gg \sigma^{2/3}),
\en
with a shock layer thickness of $L_{\rm Abel}=\tfrac{8}{3}\lambda/\sigma$ that can never go to zero. Also, we observe that the graph of eq.~\eqref{KV_aprox} is very similar, qualitatively speaking, to the one shown here in fig.~3(a). 

{\bf Remark 2:} From eq.~\eqref{KV_aprox} we find that, for sufficiently large values of $\lambda$, and a zero limit as $\xi\to +\infty$, the viscoelastic model considered in the present Letter behaves very much like the cubically perturbed Kelvin--Voigt model; see, \emph{e.g.}, Ref.~\cite{Jordan1} and those therein. 

\section{A new Burger's equation}


Let us return to  eq.~\eqref{feds} and introduce the new independent variables $\theta=\tau-a_0 \zeta$ and $s=\epsilon \zeta$, where $a_0 = \delta^{-1/2}$ and $\epsilon$ is a small parameter. If we now set $K(\zeta,\tau)=\epsilon^{1/2} \kappa(s, \theta)$ for some function $\kappa$, then  eq.~\eqref{feds} can be  approximated \emph{via} a generalized form of the modified Burger's equation~(MBE)\footnote{So named by Lee-Bapty and Crighton in 1987; see Refs.~\cite{JASA04,Jordan1} and those therein.}, namely,
\be \label{burger}
\kappa_{s}=\bar{\mu}_1 (\kappa^3)_\theta+\bar{\nu}_0\kappa_{\theta \theta}+\bar{\nu}_1(\kappa_{\theta}^3)_\theta,
\en
with the latter reducing to the former in the limit $\bar\nu_1\to 0$. Here, we have set $\bar\mu_1=\mu_1 a_0/2$, $\bar\nu_0=\nu_0 a_0/2$, and $\bar\nu_1=\nu_1 a_0/2$; and we have also assumed that $\nu_0=\mathcal{O}(\epsilon)$. Additionally, we note that when $\nu_0=0$, the structure of the travelling wave solutions of eqs.~\eqref{feds} and \eqref{burger} is exactly the same. 
We should also point out that, although eq.~\eqref{burger} is a second order differential equation, \emph{the compact kink is not an acceleration wave}, because the jump in the second derivative is not finite.
  
Moreover, because eq.~\eqref{burger} is an evolution equation, which is simpler   than the wave equation given in eq.~\eqref{feds}, other reductions to ordinary differential equations are easily found. 
For example, when $\bar{\nu}_0=0$,  eq.~\eqref{burger} admits solutions in the separable form $\kappa(\theta,s)=\psi(\theta)\phi(s)$, say. 
If $\bar{\nu}_0=0$ and $\bar{\mu}_1=0$, then eqn.~\eqref{burger} is a \textit{degenerate} diffusion equation and the separable solutions are quite simple, with an interesting structure. 
Indeed, we have 
\be
\kappa(\theta, s)=\frac{\gamma_1(\theta-\gamma_2)^2}{\sqrt{2 \gamma(\gamma_3-s)}}.
\en
Here, $\gamma$ is the separation constant and $\gamma_1$, $\gamma_2$, $\gamma_3$  are integration constants such that $24 \gamma_1^2\bar{\nu}_1-\gamma=0$, and therefore $\gamma, \gamma_3>0$. 
Clearly, these solutions have a sharp front $s=\gamma_2$; and they blow-up in space for $s=\gamma_3$.

It is interesting to consider the case of harmonic excitation and the corresponding third-harmonic order generation. Given the initial condition $\kappa(0,\theta)=\kappa_0 \cos(\omega \theta)$, where $\kappa_0$ is a positive constant, we assume the solution is given by the sum of the fundamental and third-harmonic components; i.e.,  $\kappa=\kappa_1+\kappa_3$, where
\begin{align} \notag
& \kappa_1(s,\theta)={\rm Re}\{\hat{\kappa}_1(s)\exp(\ii \omega \theta)\}=\tfrac{1}{2}\hat{\kappa}_1(s)\exp(\ii \omega \theta)+\textrm{c.c.}, \\  \notag
& \kappa_3(s,\theta)={\rm Re}\{\hat{\kappa}_3(s)\exp(3\ii \omega \theta)\}=\tfrac{1}{2}\hat{\kappa}_3(s)\exp(3\ii \omega \theta)+\textrm{c.c.}.
\end{align}
Here, ``c.c.'' denotes the complex conjugate of the proceeding term and we require that  $\left|\kappa_3\right|\ll \left|\kappa_1\right|$. By successive approximation, we have 
\be
\frac{d\hat{\kappa}_1}{ds}+\alpha \hat{\kappa}_1=0,
\en
where $\alpha=\omega^2 \bar{\nu}_0$, and
\be
\frac{d\hat{\kappa}_3}{ds}+9 \alpha \hat{\kappa}_3=-\tfrac{3}{4} \omega \hat{\kappa}_1^3(\ii \bar{\mu}_1-\omega^3 \bar{\nu}_1),
\en
where the latter is solved subject to $\kappa_3(0,\theta)=0$. 


Solving these ODEs in sequence using one of the many standard methods, we find, in turn, that
\begin{multline}
\kappa_1 = \kappa_0 \ee^{-\alpha s}\cos(\omega \theta),   \\
\kappa_3 = \dfrac{\kappa_{0}^3}{8\alpha} (\ee^{-3 \alpha s}-\ee^{-9\alpha s}) \qquad \qquad \qquad \qquad \qquad \qquad \\
\times[\bar\mu_1 \omega\sin(3 \omega \theta) + \bar\nu_1 \omega^4
\cos(3 \omega \theta)].
\label{sol:k3}
\end{multline}
These solutions reveal that the experimental measurement of the third
harmonic in soft solids with shear-dependent viscosity gives direct access
to the non-linear shear wave elastic parameter $\bar{\mu}_1$ and
non-linear dissipation parameter $\bar{\nu}_1$.
The former is measured in the low-frequency regime, whilst the latter
becomes dominant in the high-frequency regime.
 
{\bf Remark 3:}  If we let $\bar\nu_1\to 0$, and make the associations $\bar\mu_1   \mapsto  -\tfrac{1}{3}c^{-3}\beta$ and $\bar{\nu}_0 \mapsto  \delta$, then $\kappa_1$ and $\kappa_3$  respectively reduce to $v_1$ and $v_3$, which correspond to the MBE, given in \cite[eq.~(42)]{JASA04}.

{\bf Remark 4:} In attempting to satisfy $\left|\kappa_3\right|\ll \left|\kappa_1\right|$, it is helpful to know that
\be
\underset{\theta>0}{{\rm max}}|\kappa_3|\leq \frac{\kappa_{0}^3}{12\alpha\sqrt{3}}\left\{\bar\nu_1\omega^4 +|\bar\mu_1|\omega\right\},
\en
where
\be
\underset{\theta>0}{{\rm max}}|\kappa_3|=|\kappa_3|\Big{|}_{\theta=\theta^*}=\frac{\kappa_{0}^3|\bar\nu_1 \omega^4 \cos(3 \omega s)+\bar\mu_1 \omega\sin(3 \omega s)|}{12\alpha\sqrt{3}},
\en
and where $\theta^*=(6\alpha)^{-1}\ln(3)$.

{\bf Remark 5:} The inequality $\underset{\theta>0}{{\rm max}}|\kappa_3|<\underset{\theta>0}{{\rm sup}}|\kappa_1|(=\kappa_0)$ is satisfied for $\forall\,\omega\in(0, \omega^{+})$, where  $\omega^{+}$ is the only positive root of  $\bar\nu_1\omega^4 +|\bar\mu_1|\omega-12\kappa_{0}^{-2}\alpha\sqrt{3}=0$.

\section{Concluding remarks}

We have provided a clear constitutive framework that originates fields
equations admitting compact kinks.  
The constitutive framework  is relevant because it is a natural model for nonlinear 
viscoelasticity. The mechanism of compactification is clearly related to the nonlinear viscosity term. We have also derived a new Burger's equation which maintains the compactification features of the full model. The usual approximation used in nonlinear acoustics to study the generation of the third harmonic indicates that the important difference between our model and the classical one is mainly at the higher harmonics.

We established an important and original example of a how compactification may arise
in the modeling of real-world phenomena. We point out that the mechanism generating
compact waves here is completely different from those presented in \cite{Saccomandi, Destrade, Goriely},
where the compactification is made possible by an interplay between nonlinearity and
dispersion, and in \cite{Gaeta}, where compactification is made possible by considering
non-smooth potentials.

\section*{Acknowledgments}
M.D.\ is supported by a Senior Marie Curie Fellowship awarded by the European Commission (FP7). P.M.J.\ is supported by ONR/NRL funding~(PE 061153N). G.\ Saccomandi is partially supported by GNFM of INDAM.

\end{document}